\documentstyle[aps,12pt]{revtex}

\begin{document}
\author{E.N. Glass\thanks{%
Permanent address: Physics Department, University of Windsor, Ontario N9B
3P4, Canada} and J.P. Krisch}
\address{Department of Physics, University of Michigan, Ann Arbor, Michgan 48109}
\date{15 April 1999}
\title{Scale Symmetries of Spherical String Fluids }
\maketitle

\begin{abstract}
\\\\We consider homothetic maps in a family of spherical relativistic star
models. A generalization of Vaidya's radiating metric provides a fluid
atmosphere of radiation and strings. The similarity structure of the string
fluid is investigated.\\\\PACS numbers: 04.20.Jb, 04.70.Dy, 11.27.+d\newpage%
\ 
\end{abstract}

\section{INTRODUCTION}

Metric symmetries have always played a large role in the development of
exact solutions to the Einstein field equations. Often a choice of metric
symmetry is made based on an assumed symmetry of the matter distribution,
i.e. spherical symmetry for astrophysical objects or cylindrical symmetry
for a simple string$^{\cite{hiscock}}$. A homothetic motion (homothety)
describes the symmetry of scale transformations, and homothetic symmetry has
been called ''similarity of the first kind'' by Cahill and Taub$^{\cite
{cah-taub}}$. One must distinguish between {\em geometrical} and {\em %
physical} self-similarity. Geometrical similarity is a property of the
spacetime metric, whereas physical similarity is a property of the matter
fields. These need not be equivalent and the relationship between them also
depends on the nature of the matter. Yavuz and Yilmaz$^{\cite{y-y}}$
recently investigated inheritance symmetries wherein the stress energy
inherits metric symmetries of the type 
\[
{\cal L}_\xi g_{ab}=2\Psi g_{ab} 
\]
where ${\cal L}_\xi $ is the Lie derivative along the vector $\xi .$ Some of
the possibilities are 
\begin{eqnarray*}
\Psi &=&\Psi (x^a)\text{, }\xi \text{ is a conformal Killing vector,} \\
\Psi &=&1,\ \ \xi \text{ is a homothetic vector,} \\
\Psi &=&0,\ \ \xi \text{ is a Killing vector.}
\end{eqnarray*}
Carter and Henriksen$^{\cite{car-hen}}$ have introduced the idea of {\em %
kinematic self-similarity} in the context of relativistic fluid mechanics
and an extended analysis has been given by Coley$^{\cite{coley}}$. A
kinematic self-similarity vector satisfies the conditions 
\begin{eqnarray*}
{\cal L}_\xi u_a &=&const\ u_a, \\
{\cal L}_\xi h_{ab} &=&2h_{ab},
\end{eqnarray*}
where $h_{ab}=g_{ab}-u_au_b$ is the first fundamental form of the
three-spaces orthogonal to $u^a.$ The case $const\neq 1$ is called
''similarity of the second kind''.

In this work we apply the ideas of scaling and homothety to a string fluid
atmosphere. Since our primary interest is in the extended Schwarzschild mass
function $m(u,r)$ and the related string atmosphere, we apply scaling to the
mass in two different ways. First, we assume diffusive mass transport and
investigate the symmetries of the diffusion equation and second, we
investigate the scaling properties of the metric and from those derive mass
transport equations.

In the next section we briefly describe the Schwarzschild string fluid
atmosphere. The third section studies the symmetry map which takes the
diffusion equation to an ordinary differential equation. New diffusion
solutions are found. Geometric symmetries, homothetic and conformal, are
developed in section four. Mass transport is discussed in the fifth section.
One of the results of the homothetic analysis are new self-similar solutions
to the Einstein equations.

Our sign conventions are $2A_{c;[ab]}=A_eR_{\ \ cab}^e,$ and $R_{ab}=R_{\
abe}^e.$ Latin indices range over ($0$,$1$,$2$,$3$) = ($u,r,\vartheta
,\varphi $). Overdots abbreviate $\partial /\partial u$, and primes
abbreviate $\partial /\partial r$. Overhead carets denote unit vectors. We
use units where $G=c=1$. Einstein's field equations are $G_{ab}=-8\pi
T_{ab}, $ and the metric signature is (+,-,-,-).

\section{STRING FLUID ATMOSPHERE}

Recently, Glass and Krisch$^{\cite{ed-jean},\cite{ed-jean2}}$ showed that
there can be a spherically symmetric string fluid atmosphere outside a
Schwarzschild horizon. The spacetime metric is 
\begin{equation}
ds_{GK}^2=A\ du^2+2dudr-r^2(d\vartheta ^2+\text{sin}^2\vartheta d\varphi ^2).
\label{gk_met}
\end{equation}
where $A=1-2m(u,r)/r$. Initially $m(u,r)=m_0$ provides the vacuum
Schwarzschild solution in the region $r>2m_0$. The metric can be written in
a natural basis as 
\begin{equation}
g_{ab}^{GK}=\hat{v}_a\hat{v}_b-\hat{r}_a\hat{r}_b-\hat{\vartheta}_a\hat{%
\vartheta}_b-\hat{\varphi}_a\hat{\varphi}_b  \label{gk_tet}
\end{equation}
where the unit vectors are defined by 
\begin{mathletters}
\begin{eqnarray}
\hat{v}_adx^a &=&A^{1/2}du+A^{-1/2}dr,\ \ \ \ \hat{v}^a\partial
_a=A^{-1/2}\partial _u,  \label{tetv} \\
\hat{r}_adx^a &=&A^{-1/2}dr,\ \ \ \ \ \ \ \ \hat{r}^a\partial
_a=A^{-1/2}\partial _u-A^{1/2}\partial _r,  \label{tetr} \\
\hat{\vartheta}_adx^a &=&rd\vartheta ,\ \ \ \ \ \ \ \ \ \ \ \ \ \ \hat{%
\vartheta}^a\partial _a=-r^{-1}\partial _\vartheta ,  \label{tetth} \\
\hat{\varphi}_adx^a &=&r\text{sin}\vartheta d\varphi ,\ \ \ \ \ \ \ \ \hat{%
\varphi}^a\partial _a=-(r\text{sin}\vartheta )^{-1}\partial _\varphi .
\label{tetph}
\end{eqnarray}
$\hat{v}^a$ is hypersurface-orthogonal with $h_{ab}$ the first fundamental
form of the hypersurface. 
\end{mathletters}
\begin{eqnarray}
h_{ab}dx^adx^b &=&(g_{ab}^{GK}-\hat{v}_a\hat{v}_b)dx^adx^b  \label{hmet} \\
&=&-A^{-1}dr^2-r^2(d\vartheta ^2+\text{sin}^2\vartheta d\varphi ^2). 
\nonumber
\end{eqnarray}
The kinematics of the $\hat{v}^a$ flow are described by 
\begin{equation}
\hat{v}_{\ ;b}^a=a^a\hat{v}_b+\sigma _{\ b}^a-(\Theta /3)(\hat{r}^a\hat{r}_b+%
\hat{\vartheta}^a\hat{\vartheta}_b+\hat{\varphi}^a\hat{\varphi}_b),
\label{vflow}
\end{equation}
where 
\begin{mathletters}
\begin{eqnarray}
a^a &=&[\dot{m}/r+A\partial _r(m/r)]A^{-3/2}\hat{r}^a,  \label{va} \\
\sigma _{\ b}^a &=&(\Theta /3)(-2\hat{r}^a\hat{r}_b+\hat{\vartheta}^a\hat{%
\vartheta}_b+\hat{\varphi}^a\hat{\varphi}_b),  \label{vb} \\
\Theta &=&(\dot{m}/r)A^{-3/2}.  \label{vc}
\end{eqnarray}

The string distribution is described by a string bivector $\Sigma _{ab}.$
Spherical symmetry demands that the averaged string bivector will describe a
world-sheet in either the $(u,r)$ or the ($\vartheta ,\varphi )$ plane. The
string bivector is timelike and given by 
\end{mathletters}
\begin{equation}
\Sigma ^{ac}=\hat{r}^a\hat{v}^c-\hat{r}^c\hat{v}^a,  \label{stringvec}
\end{equation}
where $\Sigma ^{ac}\Sigma _c^{\ b}=\hat{v}^a\hat{v}^b-\hat{r}^a\hat{r}^b$.
The two-surfaces spanned by $\Sigma _{ab}$ are orthogonally transitive to
the two-surfaces spanned by the dual bivector 
\begin{equation}
\Sigma _{ab}^{*}=\hat{\vartheta}_a\hat{\varphi}_b-\hat{\vartheta}_b\hat{%
\varphi}_a,  \label{dual_biv}
\end{equation}
which follows from the Frobenius surface-forming condition satisfied by $%
\Sigma _{ab}$. It is also true that $\Sigma _a^{*c}\Sigma _{cb}^{*}=\hat{%
\vartheta}_a\hat{\vartheta}_b+\hat{\varphi}_a\hat{\varphi}_b$.

The Einstein tensor computed from (\ref{gk_met}) can be written as a
two-fluid system $G_{ab}^{null}+G_{ab}^{matter}$: 
\begin{eqnarray}
G_{ab} &=&(2\dot{m}/r^2)l_al_b-(2m^{\prime }/r^2)(\hat{v}_a\hat{v}_b-\hat{r}%
_a\hat{r}_b)  \label{ein_tensor} \\
&&+(m^{\prime \prime }/r)(\hat{\vartheta}_a\hat{\vartheta}_b+\hat{\varphi}_a%
\hat{\varphi}_b),  \nonumber
\end{eqnarray}
where $l_adx^a=du$. The Einstein field equations $G_{\ b;a}^a=0$ are
satisfied for arbitrary $m(u,r)$.

In Glass and Krisch$^{\cite{ed-jean},\cite{ed-jean2}}$ mass transport was
modeled by diffusion, and the diffusion equation used is given by 
\begin{equation}
\dot{\rho}={\cal D}\ r^{-2}\partial _r(r^2\partial _r\rho )  \label{rhodiffu}
\end{equation}
where ${\cal D}$ is the positive coefficient of self-diffusion (taken to be
constant).

\section{SIMILARITY MAP OF THE DIFFUSION EQUATION}

There is a similarity technique explained by Bluman and Cole$^{\cite{b-c}}$
that maps the diffusion equation into an ordinary differential equation. Our
primary interest is in the Schwarzschild mass function $m(u,r)$. New
functional solutions for $m(u,r)$ are new solutions to the field equations
for the parameter extended radiating atmosphere. The behavior of $m(u,r)$
describes the string fluid atmosphere beyond the Schwarzschild horizon
through the relations $\dot{m}=4\pi {\cal D}r^2\rho ^{\prime }$ and $4\pi
\rho =m^{\prime }/r^2$. The mass function obeys a diffusion equation 
\begin{equation}
\dot{m}={\cal D}r^2\partial _r(r^{-2}\partial _rm)  \label{m_diffu}
\end{equation}
with homogeneous solution $m_{hom}(r)=m_0+\frac 43\pi r^3\rho _0$ which can
be added to each time-dependent solution.

The similarity technique (for a fully general analysis see Bluman and Kumei$%
^{\cite{b-k}}$) requires one to introduce an independent dimensionless
variable. A standard choice in diffusion problems is the Boltzmann
transformation$^{\cite{ghez}}$: 
\begin{equation}
\eta =r\,(4{\cal D}u)^{-1/2}.  \label{eta}
\end{equation}
(Note that as a mapping from $(u,r)$ to $(u^{-1/2},\eta )$ the Jacobian is
singular implying a breakdown of the $1-1$ mapping along $r$.) The argument
of the equation, $m(u,r)$, is replaced by a dimensionless function $F(\eta )$%
. We look for a general solution of the form 
\begin{equation}
m(u,r):=c_0r^\alpha u^\beta F(\eta ).  \label{f_def}
\end{equation}
The constant $c_0$ is intended to map the dimensions of $r^\alpha u^\beta $
to mass for arbitrary constants $\alpha $ and $\beta $. Upon substituting
Eq.(\ref{f_def}) into the diffusion equation (\ref{m_diffu}) we obtain the
ordinary differential equation 
\begin{equation}
F_{\eta \eta }+2[(\alpha -1)\eta ^{-1}+\eta ]F_\eta +[\alpha (\alpha -3)\eta
^{-2}-4\beta ]F=0  \label{f_eqn}
\end{equation}
where $F_\eta $ abbreviates $dF/d\eta $.

There are many analytic solutions of Eq.(\ref{f_eqn}) which depend on the
values of $\alpha $ and $\beta $. The choice $\alpha =\beta =0$ has the
differential equation $F_{\eta \eta }+2(\eta -1/\eta )F_\eta =0$ with
solution 
\begin{equation}
F(\eta )=k_0+k_1[-\eta e^{-\eta ^2}+(\sqrt{\pi }/2)\text{erf}(\eta )],
\label{soln1}
\end{equation}
where erf$(\eta ):=(2/\sqrt{\pi })\int_0^\eta $exp($-s^2$)$ds$, lim$_{\eta
\rightarrow 0}$ erf$(\eta )=2\eta /\sqrt{\pi }$. This is the mass solution
given in Eq.(40) of Glass and Krisch$^{\cite{ed-jean2}}$ (with $k_0=0$ and
with the homogeneous solution $m_{hom}$ added). At fixed time $u$, it
describes a mass with value $m_{hom}$ + $c_0k_1$ as $\eta \rightarrow \infty 
$. At late times $c_0k_1$ is radiated away. There is no length scale in this
description so the $m_{hom}$ atmosphere is unbounded.

Other choices can be made, for example $\alpha =n,\ \ \beta =-n/2.$ This
choice has $const.\times r^nu^{-n/2}=\eta ^n$ and one can solve Eq.(\ref
{f_eqn}) or see directly from Eq.(\ref{f_def}) that 
\[
F(\eta )=\eta ^{-n}. 
\]
If we write $F(\eta )=\eta ^{-n}H(\eta )$ then $H(\eta )$ satisfies the case 
$\alpha =\beta =0$ and we have a new family of solutions parametrized by $n$%
: 
\begin{equation}
F(\eta )=\eta ^{-n}\left[ k_0+k_1[-\eta e^{-\eta ^2}+(\sqrt{\pi }/2)\text{erf%
}(\eta )]\right] .  \label{soln2}
\end{equation}
Solution (\ref{soln1}) is included here when $n=0$.

\section{SYMMETRIES}

Because the string fluid naturally lives on a two-dimensional world sheet,
the question of the symmetries of these two-dimensional subspaces is
interesting. We examine how the mass distribution and stress energy content
reflect the separate two-surface symmetries 
\begin{eqnarray}
{\cal L}_\xi (\hat{v}_a\hat{v}_b-\hat{r}_a\hat{r}_b) &=&2\mu (\hat{v}_a\hat{v%
}_b-\hat{r}_a\hat{r}_b)  \label{alph_beta_sym} \\
{\cal L}_\xi (\hat{\vartheta}_a\hat{\vartheta}_b+\hat{\varphi}_a\hat{\varphi}%
_b) &=&2\nu (\hat{\vartheta}_a\hat{\vartheta}_b+\hat{\varphi}_a\hat{\varphi}%
_b).  \nonumber
\end{eqnarray}

For similarity of the second kind, the map action must be 
\begin{eqnarray}
{\cal L}_\xi \hat{v}_a &=&\gamma \hat{v}_a,\ \ \gamma \neq 1,
\label{scnd_symm} \\
{\cal L}_\xi \hat{r}_a &=&\hat{r}_a,  \nonumber \\
{\cal L}_\xi (\hat{\vartheta}_a\hat{\vartheta}_b+\hat{\varphi}_a\hat{\varphi}%
_b) &=&2(\hat{\vartheta}_a\hat{\vartheta}_b+\hat{\varphi}_a\hat{\varphi}_b).
\nonumber
\end{eqnarray}

\subsection{Homothetic map}

The similarity vector which preserves the distinct two-surfaces of the
matter distribution in Eq.(\ref{alph_beta_sym}) is 
\begin{equation}
\xi ^a\partial _a=[\nu u_{_0}+(2\mu -\nu )u)]\partial _u+\nu r\partial _r,
\label{simvec}
\end{equation}
\ with kinematic transformations 
\begin{mathletters}
\begin{eqnarray}
{\cal L}_\xi \hat{v}_a &=&\mu \hat{v}_a,\ \ {\cal L}_\xi \hat{r}_a=\mu \hat{r%
}_a,  \label{kinv_r} \\
{\cal L}_\xi \hat{\vartheta}_a &=&\nu \hat{\vartheta}_a,\ \ {\cal L}_\xi 
\hat{\varphi}_a=\nu \hat{\varphi}_a,  \label{kinth_ph}
\end{eqnarray}
when the metric function $A$ satisfies ($\kappa :=2\mu /\nu -1$) 
\end{mathletters}
\begin{equation}
\psi \dot{A}/A+rA^{\prime }/A+\kappa -1=0.  \label{kin_constraint}
\end{equation}
with $\psi (u):=u_{_0}+\kappa u$. The constraint (\ref{kin_constraint})
requires the mass function to have the form 
\begin{equation}
r-2m(u,r)=\psi ^{\frac{2-\kappa }\kappa }f(\psi /r^\kappa )
\label{homo_mass}
\end{equation}
where $f$ is an arbitrary function.

If $\mu =\nu =\kappa =1$ then the map is homothetic with ${\cal L}_\xi
g_{ab}=2g_{ab}.$

\subsection{Another homothetic map\ }

The case $\kappa =0$ requires a separate solution. The metric function $A$
satisfies 
\begin{equation}
u_{_0}\dot{A}/A+rA^{\prime }/A=1.  \label{kin2_constraint}
\end{equation}
Constraint (\ref{kin2_constraint}) has the integral 
\begin{equation}
r-2m(u,r)=r_{_1}e^{2u/u_{_0}}\tilde{f}(e^{u/u_{_0}}r_{_0}/r)
\label{hom2_mass}
\end{equation}
with $\tilde{f}$ an arbitrary function. When $\nu =1$ and $\mu =1/2$ the $u$
dependence is eliminated from $\xi ^a$ and the transformation acts on the ($%
\vartheta ,\varphi $) two-surfaces homothetically 
\[
{\cal L}_\xi (\hat{\vartheta}_a\hat{\vartheta}_b+\hat{\varphi}_a\hat{\varphi}%
_b)=2(\hat{\vartheta}_a\hat{\vartheta}_b+\hat{\varphi}_a\hat{\varphi}_b) 
\]
but preserves the scale of the string two-surfaces 
\[
{\cal L}_\xi (\hat{v}_a\hat{v}_b-\hat{r}_a\hat{r}_b)=\hat{v}_a\hat{v}_b-\hat{%
r}_a\hat{r}_b. 
\]

\subsection{Interpreting the scale parameter}

Under the action of the homothety $\xi ^a\partial _a=(u_{_0}+u)\partial
_u+r\partial _r$ the acceleration of $\hat{v}^a$, given in Eq.(\ref{va}),
has the following Lie derivative: with $a^b=a\hat{r}^b$, $a:=[\dot{m}%
/r+A\partial _r(m/r)]A^{-3/2}$%
\begin{equation}
{\cal L}_\xi a^b=\left( \frac{a_\xi }a-1\right) a^b  \label{lie_a}
\end{equation}
where $a_\xi :=[(u_{_0}+u)\partial _u+r\partial _r]a$. Similarly the
rate-of-shear given in Eq.(\ref{vb}) and Eq.(\ref{vc}) obeys 
\begin{equation}
{\cal L}_\xi \sigma _{\ b}^a=\left( \frac{\Theta _\xi }\Theta -1\right)
\sigma _{\ b}^a  \label{lie_shear}
\end{equation}
where $\Theta _\xi =[(u_{_0}+u)\partial _u+r\partial _r]\Theta $.

There is no information to be gained by analyzing the scaling properties of
the Raychaudhuri equation 
\[
a_{;b}^b-\sigma _{ab}\sigma ^{ab}-\Theta ^2/3-\Theta _{,a}\hat{v}^a=-R_{ab}%
\hat{v}^a\hat{v}^b 
\]
since it is identically satisfied by $g_{ab}^{GK}$.

\subsection{Conformal map}

The case $\mu =\nu =\kappa =1$, with $\psi =u_{_0}+u$, has an interesting
conformal symmetry. We see from Eq.(\ref{homo_mass}) that $A=(\psi /r)f(\psi
/r)=F(\psi /r)$. Metric (\ref{gk_met}) is written as 
\begin{equation}
ds_{GK}^2=F(\psi /r)du^2+2dudr-r^2d\Omega ^2.  \label{new_1}
\end{equation}
We define a new coordinate $y:=r/\psi $ and rewrite (\ref{new_1}) as 
\begin{equation}
ds_{GK}^2=[F(1/y)+2y]du^2+2\psi dudy-y^2\psi ^2d\Omega ^2.  \label{new_2}
\end{equation}
Now we factor out $\psi ^2$ and introduce a new time coordinate $dw:=du/\psi 
$ to obtain 
\[
ds_{GK}^2=\psi ^2\left[ (F+2y)dw^2+2dwdy-y^2d\Omega ^2\right] . 
\]
Upon choosing $F(1/y)=1-2M(y)/y-2y$, $M(y)$ arbitrary, we have 
\begin{equation}
ds_{GK}^2=e^{2w}\left[ (1-2M/y)dw^2+2dwdy-y^2d\Omega ^2\right] .
\label{conf_met}
\end{equation}
The argument above shows that the similarity transformation generated by
vector $\xi ^a\partial _a=(u_{_0}+u)\partial _u+r\partial _r$ conformally
relates the radiating string atmosphere of metric (\ref{gk_met}) to a
previously identified family of static string atmospheres$^{\cite{ed-jean2}}$
i.e. 
\begin{equation}
{\cal L}_\xi g_{ab}^{GK}=2e^{2w}g_{ab}^{static}.  \label{lie_gk}
\end{equation}

\subsection{Similarity of the second kind}

For similarity vector $\xi ^a\partial _a=(u+u_{_0})\partial _u+r\partial _r$
the metric function $A$ must satisfy 
\begin{equation}
(u_{_0}+u)\dot{A}/A+rA^{\prime }/A=\gamma -1.  \label{kin3_constraint}
\end{equation}
Equation (\ref{kin3_constraint}) has solution 
\begin{equation}
r-2m(u,r)=r_{_2}(u_{_0}+u)^\gamma h[(u_{_0}+u)/r]  \label{scnd_mass}
\end{equation}
where $h$ is an arbitrary function and $\gamma \neq 1$.

\section{MASS TRANSPORT}

The mass functions found by similarity analysis obey certain transport
equations. Most of the transport equations have the form of the
''telegrapher'' equation. This can describe dispersive and lossy
electromagnetic wave motion$^{\cite{strat}}$. Some forms have been
interpreted by Kac$^{\cite{kac}}$ as a random Poisson process. Mass
transport through the atmosphere is affected by the homothetic symmetries.
The transport equations can be constructed from the similarity solutions of
the previous section.

\subsection{$\kappa =0\ \ $Homothety}

Differentiation of equation (\ref{kin2_constraint}), a constraint on the
mass function, yields an inhomogeneous wave equation 
\begin{equation}
\ddot{A}-3\dot{A}/u_{_0}-(r/u_{_0})^2\nabla ^2A=-2A/u_{_0}^2.
\label{ur-wave2}
\end{equation}

\subsection{$\kappa =1\ $\ Homothety}

Recall metric function $A=1-2m(u,r)/r$. One can see directly from Eq.(\ref
{homo_mass}) that $A=(\psi /r)f(\psi /r)=F(\psi /r)$ with $\psi =u_{_0}+u$. $%
A$, and thus $m/r$, satisfies a wave equation on the flat tangents to the $%
\hat{v}_a\hat{v}_b-\hat{r}_a\hat{r}_b$ two-spaces. Defining $\tau =$ln($\psi 
$) and $z=$ln($r$) we have 
\begin{equation}
A=F(\tau -z).  \label{tau-z}
\end{equation}
It is clear that $A$, generated by homothety $\xi ^a\partial
_a=(u_{_0}+u)\partial _u+r\partial _r$, satisfies the wave equation 
\begin{equation}
\frac{\partial ^2A}{\partial \tau ^2}-\frac{\partial ^2A}{\partial z^2}=0.
\label{tau-z-wave}
\end{equation}

Alternatively, we can find a wave equation on the curved manifold by writing 
\[
\partial F/\partial u=(1/r)\hat{F},\ \ \ \ \partial F/\partial r=-(\psi /r^2)%
\hat{F} 
\]
where $\hat{F}$ is the derivative of $F$ with respect to its argument. It
follows that 
\[
\ddot{A}=(1/r^2)\hat{\hat{F\,}},\ \ \ \ (r^2A^{\prime })^{\prime }=(\psi
^2/r^2)\hat{\hat{F}\,.}\ 
\]
Thus 
\begin{equation}
\ddot{A}-v_s^2\nabla ^2A=0  \label{ur-wave}
\end{equation}
where $\nabla ^2=r^{-2}(\partial /\partial r)r^2(\partial /\partial r)$ and $%
v_s=r/\psi $. The wave speed varies with $u$ and $r$.

If $v_s$ were constant $v$, then Eq.(\ref{ur-wave}) would have the general
solution 
\begin{equation}
A(u,r)=\frac{f(r-vu)}r+\frac{g(r+vu)}r  \label{wave_soln}
\end{equation}
in terms of two arbitrary functions $f$ and $g$. Substituting $A=f/r$ into (%
\ref{ur-wave}) one finds 
\[
(v^2-v_s^2)\hat{\hat{f}}=0. 
\]
This reflects ''damped, yet relatively undistorted, progressing wave
solutions'' \cite{cour-hil}, a special case of the telegrapher's equation.

For new time coordinate $t\mapsto $ $u+u_{_0}=e^{t/t_{_0}}$ and with $%
A_t:=\partial A/\partial t$, Eq.(\ref{ur-wave}) transforms to 
\begin{equation}
A_{tt}-\,A_t/t_{_0}-(r/t_{_0})^2\nabla ^2A=0.  \label{tr1-wave}
\end{equation}

\subsection{$\kappa \geq 2$ Two-surface symmetry}

With $A=1-2m(u,r)/r=\psi ^{\frac{2-\kappa }\kappa }r^{-1}f(\psi /r^\kappa )$
we can write 
\[
A=r^{1-\kappa }H(\psi /r^\kappa ),\ \ \ \ H:=(\psi /r^\kappa )^{\frac{%
2-\kappa }\kappa }f. 
\]
Differentiation yields 
\[
\ddot{A}=\kappa ^2r^{1-3\kappa }\hat{H}\ 
\]
and 
\[
(r^2A^{\prime })^{\prime }=(1-\kappa )(2-\kappa )r^{1-\kappa }H-3\kappa
(1-\kappa )\psi r^{1-2\kappa }\hat{H}+\kappa ^2\psi ^2\hat{H}\ . 
\]
It follows that 
\[
(r^2A^{\prime })^{\prime }=(1-\kappa )(2-\kappa )A-3(1-\kappa )\psi \dot{A}%
+\psi ^2\ddot{A}. 
\]
Transforming to a new time coordinate $e^{t/t_{_0}}=\psi =u_{_0}+\kappa u$
yields the inhomogeneous wave equation 
\begin{equation}
A_{tt}+(2-3/\kappa )\,A_t/t_{_0}-(r/t_{_0})^2\nabla ^2A=(1-1/\kappa
)(2/\kappa -1)\,A/t_{_0}^2  \label{tr-wave}
\end{equation}
where $A_t:=\partial A/\partial t.$

\subsection{Similarity of the second kind}

Differentiation of the constraint on metric function $A$, equation (\ref
{kin3_constraint}), yields the homogeneous wave equation 
\begin{equation}
\ddot{A}+v_s(\frac{\gamma -1}r)\dot{A}-v_s^2\nabla ^2A+v_s^2(\frac{\gamma -1%
}{r^2})(rA^{\prime })=0,  \label{ur-wave3}
\end{equation}
where $v_s=r/(u_{_0}+u)$ and $\gamma \neq 1$. As above, the wave speed
varies with $u$ and $r$.

\section{Discussion}

Similarity is physically important since scaling behavior may offer clues
about possible relationships between macroscopic and microscopic physics
(i.e. Ehrenfest's classical adiabatic invariants and quantization rules).
Using scaling, one can model long term behaviors with single solutions to
the field equations in which only the scaling variable changes as a function
of time. Self-similar behavior is an important aspect of many evolutionary
processes both linear and non-linear$^{\cite{ben-col}}$. The simplifications
of the non-linear field equations of general relativity are a good example
of the value of similarity methods. In addition, we have seen that the
special homothety of fluid two-surfaces can be associated with self similar
behavior in the fluid parameters.

In this paper our primary interest is in the extended Schwarzschild mass
function $m(u,r)$ and the related string atmosphere. We applied scaling to
the mass in two different ways. First, we assumed a mass transport and
investigated the scaling properties and second, we investigated the scaling
properties of the metric and from those derived mass transport equations. In
the first case, assuming diffusive mass transport with a Boltzmann scaling
variable, we developed a new family of diffusive mass functions and the
associated family of string atmospheres. In the second case, we examined the
scaling symmetries of the orthogonal two-surfaces $(u,r)$ and $(\vartheta
,\varphi )$. A 2-parameter similarity generator acted separately on the $%
(u,r)$ two-surface containing the string fluid and the orthogonal $%
(\vartheta ,\varphi )$ two-surface subject to the mass parameter obeying a
constraining first order differential equation. The similarity map affects
all metric components equally when the parameters are both equal to 1. For
this case, where the transformation is a homothety for the entire spacetime,
the mass constraint conformally relates a radiating string atmosphere and a
static atmosphere. Other parameter choices could be made, for example, the
choices which remove time dependence from the generator. This time
independent mapping acts on the $(\vartheta ,\varphi )$ two-surface
homothetically while preserving the scale of the $(u,r)$ string two-surface.
For all the parameter choices associated with the scaling action of the
generator, a mass transport equation is implied. This equation is, in
general, the telegrapher's equation. The telegrapher's equation and the
diffusion equation have both macroscopic and microscopic interpretations$^{%
\cite{kac},\cite{nelson1},\cite{nelson2}}$. The appearance of both of these
mass transport equations in conjunction with the description of a
macroscopic string fluid atmosphere is suggestive of the quantum nature of
the fundamental string fluid bits. The classical continuum fluid describes
only the averaged fluid behavior, with the mass transport equations
suggesting the underlying quantum nature of the fluid.

\end{document}